\documentclass[aps,prx,twocolumn,superscriptaddress, nofootinbib,  longbibliography]{revtex4-2}
\usepackage{graphicx} 
\usepackage{dcolumn}  
\usepackage{bm}    
\usepackage{amssymb}  
\usepackage{amsfonts, amsmath, amssymb, mathtools}
\usepackage{amsmath}

\usepackage{lipsum}
\usepackage[normalem]{ulem}
\usepackage{commath}
\usepackage{xcolor}
\usepackage{scalerel}[2016-12-29]

\usepackage[binary-units=true]{siunitx}
\usepackage{hyperref}
\usepackage{cleveref}
\RequirePackage{textcase}
\DeclareUnicodeCharacter{2009}{\,}

\makeatletter
\def\maketitle{
\@author@finish
\title@column\titleblock@produce
\suppressfloats[t]}
\makeatother

\begin{document}

\newcommand{\condProb}{\operatorname{P}\expectarg}
\DeclarePairedDelimiterX{\expectarg}[1]{(}{)}{%
 \ifnum\currentgrouptype=16 \else\begingroup\fi
 \activatebar#1
 \ifnum\currentgrouptype=16 \else\endgroup\fi
}

\newcommand{\innermid}{\nonscript\;\delimsize\vert\nonscript\;}
\newcommand{\activatebar}{%
 \begingroup\lccode`\~=`\|
 \lowercase{\endgroup\let~}\innermid
 \mathcode`|=\string"8000
}

\newcommand{\err}{\epsilon} 
\newcommand{\bra}[1]{\left<#1\right|}
\newcommand{\ket}[1]{\left|#1\right>}
\newcommand{\bket}[2]{\left<#1~|~#2\right>}
\newcommand{\tr}[1]{\text{Tr}\left(#1\right)}
\newcommand{\kket}[1]{\left|\left|#1\right>\right>}
\newcommand{\bbra}[1]{\left<\left<#1\right|\right|}
\newcommand{\ba}{\boldsymbol{a}}
\newcommand{\bb}{\boldsymbol{b}}
\newcommand{\bc}{\boldsymbol{c}}
\newcommand{\bd}{\boldsymbol{d}}
\newcommand{\bh}{\boldsymbol{h}}
\newcommand{\bq}{\boldsymbol{q}}
\newcommand{\bp}{\boldsymbol{p}}
\newcommand{\bQ}{\boldsymbol{Q}}
\newcommand{\bP}{\boldsymbol{P}}
\newcommand{\bE}{\boldsymbol{E}}
\newcommand{\mE}{\mathcal{E}}
\newcommand{\Tr}{\text{Tr}}
\renewcommand{\Re}{\text{Re}}
\renewcommand{\Im}{\text{Im}}
\newcommand{\ta}{\tilde{\alpha}}
\newcommand{\bO}{\boldsymbol{\mathcal{O}}}
\newcommand{\br}{\boldsymbol{r}}
\newcommand{\bR}{\boldsymbol{R}}
\newcommand{\bK}{\boldsymbol{K}}
\newcommand{\bJ}{\boldsymbol{J}}
\newcommand{\bH}{\boldsymbol{H}}
\newcommand{\bU}{\boldsymbol{U}}
\newcommand{\bM}{\boldsymbol{M}}
\newcommand{\bX}{\boldsymbol{X}}
\newcommand{\bZ}{\boldsymbol{Z}}
\newcommand{\bY}{\boldsymbol{Y}}
\newcommand{\bI}{\boldsymbol{I}}
\newcommand{\bL}{\boldsymbol{L}}
\newcommand{\bT}{\boldsymbol{T}}
\newcommand{\bD}{\boldsymbol{D}}
\newcommand{\bn}{\boldsymbol{n}}
\newcommand{\bS}{\boldsymbol{S}}
\newcommand{\bsigma}{\boldsymbol{\sigma}}
\newcommand{\bSigma}{\boldsymbol{\Sigma}}
\newcommand{\bDelta}{\boldsymbol{\Delta}}
\newcommand{\bPi}{\boldsymbol{\Pi}}
\newcommand{\red}[1]{\textcolor{red}{#1}}
\newcommand{\green}[1]{\textcolor{green}{#1}}
\newcommand{\blue}[1]{\textcolor{blue}{#1}}
\newcommand{\bphi}{\boldsymbol{\varphi}}
\newcommand{\NN}{\mathcal N}

\raggedbottom

\renewcommand{\thesubsection}{\thesection.\arabic{subsection}}
\renewcommand{\thesubsubsection}{\thesubsection.\arabic{subsubsection}}

\title{Protecting the quantum interference of cat states by phase-space compression
}

\author{Xiaozhou Pan}
\email[Corresponding author: ]{xiaozhou@nus.edu.sg}
\author{Jonathan Schwinger}
\email[Corresponding author: ]{jschwinger@u.nus.edu}
\author{Ni-Ni Huang}
\author{Pengtao Song}
\affiliation{Centre for Quantum Technologies, National University of Singapore, Singapore}
\author{Weipin Chua}
\affiliation{Department of Physics, National University of Singapore, Singapore}
\author{Fumiya Hanamura}
\affiliation{Department of Applied Physics, School of Engineering, The University of Tokyo, Japan}
\author{Atharv Joshi}
\author{Fernando Valadares}
\affiliation{Centre for Quantum Technologies, National University of Singapore, Singapore}
\author{Radim Filip}
\affiliation{Department of Optics, Palacky University, Czech Republic}
\author{Yvonne Y. Gao}
\email[Corresponding author: ]{yvonne.gao@nus.edu.sg}
\affiliation{Centre for Quantum Technologies, National University of Singapore, Singapore}
\affiliation{Department of Physics, National University of Singapore, Singapore}
\date{\today}

\begin{abstract}
Cat states, with their unique phase-space interference properties, are ideal candidates for understanding fundamental principles of quantum mechanics and performing key quantum information processing tasks. However, they are highly susceptible to photon loss, which inevitably diminishes their quantum non-Gaussian features. Here, we protect these non-Gaussian features against photon loss by compressing the phase-space distribution of a cat state. We achieve this compression with a deterministic technique based on the echo conditional displacement operation in a circuit QED device. We present a versatile technique for creating robust non-Gaussian continuous-variable resource states in a highly linear bosonic mode and manipulating their phase-space distribution to achieve enhanced resilience against photon loss. Compressed cat states offer an attractive avenue for obtaining new insights into quantum foundations and quantum metrology, and for developing inherently more protected bosonic codewords for quantum error correction.
\end{abstract}

\maketitle

Schrödinger's cat state is a superposition of macroscopically distinct quantum states, famous from the iconic Gedankenexperiment of a cat being simultaneously dead and alive~\cite{schrodinger1935_gegenwartige}. This concept morphed into the simplified version of a superposition of two sufficiently large (i.e., negligibly overlapping) coherent states with opposite phases, i.e., $|\alpha\rangle\pm|-\alpha\rangle$. Such large cat states are of great interest because they exhibit unique non-classical attributes such as sub-Planck phase-space structures~\cite{zurek2001_sub}
and non-Gaussian interference features~\cite{schleich1991_nonclassical, gerry1997_quantum}. Apart from being a gateway to better understand the fundamental physics of quantum decoherence~\cite{raimond2006_exploring}, cat states also act as the backbone for continuous-variable (CV) quantum information processing, including quantum metrology~\cite{munro2002weak, joo2011quantum, facon2016_sensitive, knott2016practical, duivenvoorden2017_single}, quantum teleportation and cryptography~\cite{van2001_entangled, jeong2001_quantum, 
sangouard2010_quantum, brask2010_hybrid, lee2013_near}, and the development of error-correcting codes for fault-tolerant quantum computing~\cite{ralph2003_quantum, lund2008_fault} by tracking the photon parity as the error syndrome in a hardware-efficient manner~\cite{leghtas2013_hardware, mirrahimi2014_dynamically, leghtas2015_confining, ofek2016_extending, puri2017_engineering, touzard2018_coherent, puri2019_stabilized, guillaud2019_repetition, grimm2020_stabilization, puri2020_bias, lescanne2020_exponential, chamberland2022_building}.

The appeals of the cat state's distinctive quantum non-Gaussian interference characteristics for probing fundamental physics and CV quantum information applications have motivated significant experimental efforts to create and manipulate them in various hardware platforms, such as optical systems~\cite{ourjoumtsev2009_preparation, huang2015optical, asavanant2017_generation}, vibrational states of a trapped ion~\cite{wineland2013_nobel, kienzler2016_observation}, Rydberg atoms~\cite{haroche2013_nobel, deleglise2008_reconstruction}, and circuit quantum electrodynamics (cQED)~\cite{vlastakis2013_deterministically, wang2016schrodinger}. In particular, the cQED platform in the form of 3D superconducting cavities controlled by one or more non-linear ancillary modes has enabled the creation~\cite{vlastakis2013_deterministically} as well as universal control~\cite{heeres2015_cavity,krastanov2015_universal} of large and highly coherent cat states, making it an excellent candidate for storing and manipulating quantum information and non-Gaussian resource states~\cite{krantz2019quantum, kjaergaard2020_superconducting, blais2021_circuit}. However, the desired interference features in cat states, formed by the coherent superposition of $\pm|\alpha\rangle$, are notoriously delicate. Their rapid decay under photon loss, the dominant error channel of the cavity mode, directly signifies the loss of quantum non-Gaussian characteristics~\cite{leggett1987_dynamics, caldeira1983_quantum,caldeira1983_path, caldeira1985_influence, leggett1984_schrodinger, zurek1981_pointer, moore2012_frontiers, zurek1991_quantum, kim1992_schrodinger}. 

Preservation of these quantum interference features can be deterministically achieved by compressing the state in phase space. Using the characteristic function representation, which describes the spectral landscape of the quantum states, photon loss can be modeled as a point-wise scaling and multiplication with a low-pass Gaussian filter~\cite{leonhardt2010_Essential}. For large cat states, the interference features are located far away from the origin in characteristic function as they contain higher frequency components. Their amplitudes, therefore, diminish significantly under the action of the filter, shown in Fig.~\ref{fig:concept}(a), making these features highly vulnerable to photon loss. Phase-space compression, shown in Fig.~\ref{fig:concept}(b), deterministically reshapes the characteristic function landscape such that these higher frequency components can fit within the low-pass filter. Thus, the interference features of the compressed cat state are intrinsically more protected from the action of the Gaussian filtering due to photon loss~\cite{filip2001_amplification, serafini2004_minimum, menzies2009_gaussian, filip2013gaussian, Le2018_slowing, park2022_slowing, teh2020_overcoming}. While the creation of compressed cat states has been shown in quantum optics~\cite{etesse2015_experimental, huang2015_optical} and ion trap devices~\cite{lo2015_spin, fluhmann2020_direct}, their time dynamics and resilience against photon loss have only been experimentally probed using an optical parametric process~\cite{Le2018_slowing}. This is due to the challenge of implementing fast non-linear controls to create cat states and perform phase-space compression on them without introducing excessive non-linearity in the quantum harmonic oscillator.

\begin{figure}[t!]
\centering
\includegraphics[width=\columnwidth]{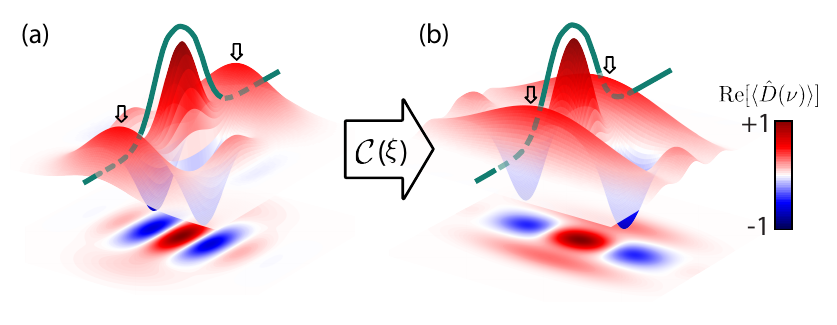}
\caption{\textbf{Protection of quantum interference against photon loss.} (a) Cat states created in superconducting cavities suffer predominantly from photon loss, which acts as a symmetric low-pass Gaussian filter (green) with width determined by $\sim1/\kappa t$ in the characteristic function representation,  where $\kappa$ is the rate of photon loss.  The quantum interference features, represented by the amplitude of the blobs (black arrows), diminish substantially over time as they are composed of higher frequency components, which are more susceptible to the low-pass filter imposed by photon loss. (b) Preservation of the quantum non-Gaussianity can be achieved by compressing the phase space of the cat states, symbolized by the arrow with the compression coefficient $\xi$. Under the action of the compression operation, the interference blobs are pushed closer to the origin. This makes the resulting state significantly less susceptible to the  filter and effectively preserves its quantum interference features against photon loss.}
\label{fig:concept}
\end{figure}

In this work, we demonstrate the deterministic protection of the quantum non-Gaussian interference of cat states by engineering their phase-space distribution to be more compact. We realize this compression using a versatile conditional displacement operation, which employs only native single cavity and transmon gates in a standard cQED device. We directly probe the time dynamics of quantum coherence using features in the characteristic function, and demonstrate enhanced resilience of the compressed cat's quantum non-Gaussian interference in the presence of intrinsic losses within the hardware. Our technique offers a highly versatile operation for the creation and storage of CV resource states in bosonic modes with minimal anharmonicity. More generally, our study illustrates a powerful framework for enhancing the noise-resilience of bosonic quantum states via the universal features of the echo conditional displacement gates~\cite{Eick2021_fast} to engineer and reshape their phase space to be more optimal against the local loss mechanisms~\cite{park2022_slowing}. Furthermore, our realization of the compressed cat states provides an important new ingredient to implement more intrinsically-protected logical codewords for fault-tolerant quantum computing~\cite{mirrahimi2014_dynamically, joshi2021_quantum, xu2022_autonomous, hillmann2022_quantum} and a promising candidate for quantum metrology~\cite{munro2002weak, joo2011quantum, facon2016_sensitive, knott2016practical, duivenvoorden2017_single}.

Our experimental setup is a three-dimensional (3D) cQED architecture, where a 3D cavity couples to a planar chip containing an ancillary transmon and a low-Q readout resonator, as shown in Fig.~\ref{fig:sqvac}(a). The 3D cavity, machined out of high-purity (4N) aluminium,  provides a high-Q bosonic mode for the creation and storage of CV quantum states. The tasks of storing large cat states and performing phase-space compression on them over time require the cavity to inherit minimal anharmonicity from the transmon. As such states occupy a large span of energy levels in the harmonic oscillator, they suffer from significant distortions in phase space even in the presence of moderate non-linearity in the cavity. Therefore, we design the system to have a weak dispersive coupling, with $\chi / 2\pi\approx 40$\,kHz. This effectively suppresses the inherited non-linearity on the bosonic mode to $K / 2\pi \approx 10$\,Hz. Considering the negligible cavity non-linearity, our system is well described by the dispersive Hamiltonian
$\mathbf{\hat{H}(t)}/\hbar=-\frac{\chi}{2}\mathbf{\hat{a}}^{\dagger} \mathbf{\hat{a}}\sigma_{\mathbf{z}},$ where $\mathbf{\hat{a}}$ and $\mathbf{\hat{a}}^{\dagger}$ are the annihilation and creation operators of the cavity mode and $\sigma_{\mathbf{z}}$ the Pauli Z operator of the transmon. 

While these device parameters ensure that our cavity is a good harmonic oscillator, it comes with the cost of reduced controllability as we typically rely on $\chi$ to perform universal gates on the bosonic mode~\cite{krastanov2015_universal, heeres2017_implementing}. Here, we sidestep this limitation by employing a technique called echo conditional displacement (ECD)~\cite{Camp2020_quantum, Eick2021_fast}. The ECD gate takes advantage of large cavity displacements $\alpha_0$, which act as an extended lever in phase space such that a phase that is dependent on the transmon state can be rapidly accumulated despite the small $\chi$. This method allows conditional displacements operations to be enacted in a duration $\propto \frac{1}{\chi \alpha_0}$ instead of $\propto \frac{1}{\chi}$, and thus, making it possible to balance the need for minimising non-linearity and effect universal control on the cavity~\cite{Eick2021_fast}. 

Practically, the ECD gate can be readily realized on standard cQED devices. The operation only requires single transmon rotations and unconditional cavity displacements. The echo arises from a $\pi$-pulse played in the middle of the sequence to suppress low frequency noise on the transmon. Following the technique introduced in Refs.~\cite{Camp2020_quantum, Eick2021_fast}, we can enact the following effective Hamiltonian,
\begin{equation}\label{eq:ecd_hamiltonian}
     \frac{\mathbf{\hat{H}}(t)}{\hbar}=-\frac{\chi}{2}\left(\alpha(t) \mathbf{\hat{a}}^{\dagger}+\alpha^*(t) \mathbf{\hat{a}}\right) \sigma_{\mathbf{z}},
\end{equation} 
where $\alpha(t)$ is the cavity's classical response in the presence of the resonant displacement pulses. Under Eq.~\ref{eq:ecd_hamiltonian}, an ECD unitary operation, given by 
\begin{equation}
    \mathrm{ECD}(\alpha) =\hat{D}(\alpha / 2)|g\rangle\langle e|+\hat{D}(-\alpha / 2)| e\rangle\langle g|,
\end{equation}
is enacted on the cavity-transmon system, such that the phase of the cavity displacement is dependent on the transmon state. In our device, we can perform an ECD gate with $\alpha = 1$ in $688\,$ns, which compares favourably to the cavity's single-photon lifetime of $T^c_1=260\,\mu$s and dephasing timescale of $T^c_{\phi}\approx$ 5\,ms, and with the transmon's coherence properties of $T_1=18\,\mu$s and $T_{2e}=20\,\mu$s~\cite{pan2022_supp}.

\begin{figure}[t!]
\includegraphics[scale=1]{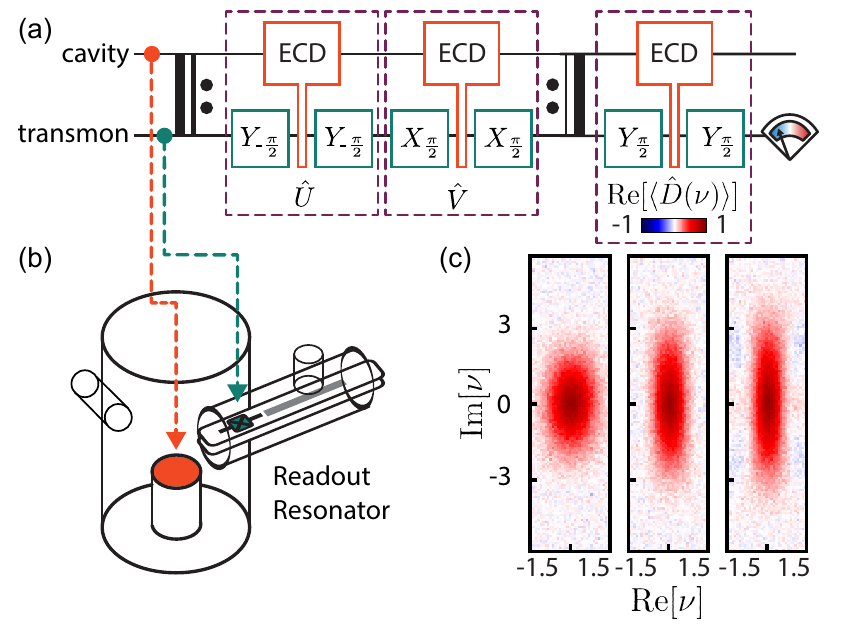}
\caption{\textbf{Deterministic phase-space compression in cQED} (a) With vacuum as the initial state, the protocol generates compressed vacuum states based on Ref.~\cite{Hastrup2021_Unconditional}. The $\hat{U}$ and $\hat{V}$ operations are each decomposed into two transmon rotations and one ECD gate. The resulting states are measured by their characteristic functions, which are performed using the ECD gates. 
(b) Color coded schematic of the device showing the storage cavity, transmon and readout mode.
(c) Real part of the characteristic functions compressed vacuum states generated by 3 repetitions of the $\hat{U}\hat{V}$ operations with phase-space compression of $\text{-}3$\,dB, $\text{-}6.7$\,dB, and $\text{-}7.6$\,dB along $\mathrm{Re}[\nu]$, respectively.}
\label{fig:sqvac}
\end{figure}

Building upon the ECD gate, we implement a deterministic compression protocol to generate compressed vacuum states with the technique proposed in Ref.~\cite{Hastrup2021_Unconditional}. This procedure relies on two unitary operations $\hat{U}_k$ and $\hat{V}_k$, where $\hat{U}_k$ = exp$(iu_k\hat{P}\hat{\sigma}_x)$ and $\hat{V}_k$ = exp$(iv_k\hat{X}\hat{\sigma}_y)$ with $\hat{X} = (\hat{a} + \hat{a}^\dagger)/2$ and $\hat{P} = i(\hat{a}^\dagger - \hat{a})/2$. Conceptually, $\hat{U}_k$ displaces the cavity by an amount proportional to $u_k$ in opposite directions in phase space depending on the transmon state, while $\hat{V}_k$ approximately disentangles cavity and transmon. Then, one set of $\hat{U}\hat{V}$ brings the initial vacuum state into two coherent states with opposite phases. Due to the interference of these two components, the resulting cavity state is compressed in one quadrature and elongated in the other. The $\hat{U}$ and $\hat{V}$ operations can be readily implemented experimentally in our system through the following decomposition,
\begin{align}
     \hat{U}_k &= R_y(-\frac{\pi}{2})\mathrm{ECD}(u_k)R_y(-\frac{\pi}{2}),\\
    \hat{V}_k &= R_x(\frac{\pi}{2})\mathrm{ECD}(iv_k)R_x(\frac{\pi}{2}),
\end{align}
featuring ECD gates and numerically optimised coefficients $u_k$ and $v_k$. Repeating them with the appropriate coefficients allows us to create a superposition of $2^N$ coherent states with a Gaussian distribution, which is effectively an approximate squeezed vacuum state (Fig.~\ref{fig:sqvac}(a)). This procedure mimics the action of the squeezing used in the 
previous study using an optical parametric process~\cite{Le2018_slowing}. However, our gate-based technique offers enhanced versatilility and can be extended to achieve other forms of phase-space engineering afforded by the universality of the ECD operations~\cite{Eick2021_fast}. 


Based on the above strategy, we generate three vacuum states with $\text{-}3$\,dB, $\text{-}6.7$\,dB, and $\text{-}7.6$\,dB of compression in one quadrature. To achieve this, we use three steps of the $\hat{U}\hat{V}$ operations and optimize the interaction parameters, $u_k$ and $v_k$, for maximum overlap with the ideal squeezed state~\cite{pan2022_supp}. We then employ the ECD gate to perform characteristic function tomography~\cite{Camp2020_quantum}. The real part of the measured characteristic functions, are shown in Fig.~\ref{fig:sqvac}(c). Theoretically, each additional step of $\hat{U}\hat{V}$ increases the possible degree of compression by 3-4\,dB. However, practically, the transmon's $T_{2e}\sim20\,\mu$s imposes a limit of three repetitions, which nonetheless allows us to achieve up to $\text{-}7.6$\,dB reduction in the width of the Gaussian distribution in the $\mathrm{Re}[\nu]$ quadrature in $\sim4\,\mu$s. While the deviation from ideal squeezing is apparent, it does not undermine the protocol's effectiveness in compressing the characteristic function of the vacuum state. 

\begin{figure}[t!]
\includegraphics[scale=1]{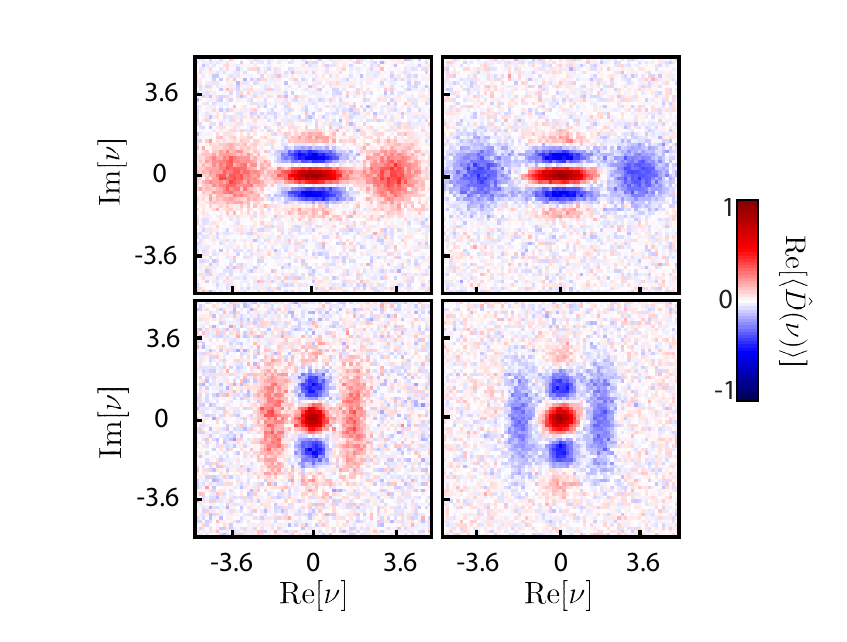}
\caption{
\textbf{Characteristic functions of cat and compressed cat states.} 
Real part of the measured characteristic function of even and odd cat states of size $\alpha=1.8$ with 0\,dB (top) and approximately $\text{-}6.7$\,dB (bottom) compression. The blobs, which correspond to quantum interference, are centered at $2\alpha \approx 3.6$ and $2\alpha \approx 1.8$, respectively. They also reflect the photon number parity of the state in the characteristic function representation, and are pushed closer to the origin by compressing the state. } 
\label{fig:cat_sqcat} 
\end{figure}

\begin{figure*}[bht!]
\includegraphics[scale=1]{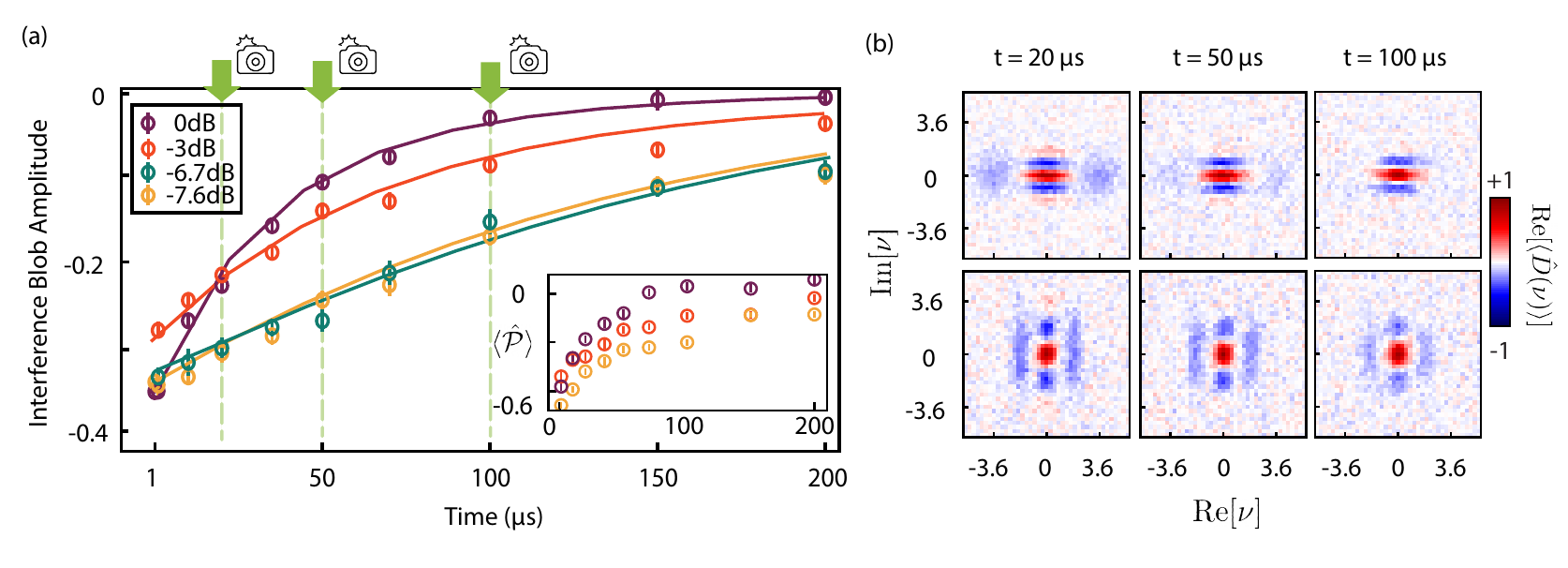}
\caption{\textbf{Time dynamics of interference features in the presence of intrinsic losses in the hardware.} (a) The measured contrast (marker) of the blobs in the characteristic functions of odd cat states with 0, $\text{-}3$, $\text{-}6.7$, and $\text{-}7.6$\,dB of compression along $\mathrm{Re}[\nu]$. We find time decay constants of $42\pm 2\,\mu$s, $87 \pm 5\,\mu$s, $145 \pm 6\,\mu$s, and $147 \pm 8\,\mu$s, respectively. The decay is significantly slower for the compressed cat states, demonstrating that compression and the resulting compactness in phase space protects the quantum interference features in cat states.  They agree closely with the simulated behavior (solid lines) with cavity photon loss as the dominant error. Inset: the equivalent decay of the photon number parity, $\langle\hat{\mathcal{P}}\rangle$, extracted from the Wigner functions of the same set of states~\cite{pan2022_supp}. The preservation of the odd parity indicates that compressed cat states are more resilient towards photon loss, in agreement with the metric obtained using characteristic functions. All associated uncertainties are extracted by standard bootstrapping techniques. (b) Selected characteristic functions at different decay times, showing the effective preservation of the blobs due to compression.}
\label{fig:time_dynamics}
\end{figure*}

Using these compressed vacuum states as the input, we create compressed cat states using an additional $\hat{V}$ operation, followed by a single-shot measurement of the transmon. The $\hat{V}$ gate conditionally displaces the compressed cavity state in opposite directions in phase space. As a result, the transmon's ground state is entangled with an  compressed even cat state and its excited state entangled with the odd one. We then utilize the single-shot measurement to project the transmon state before performing the characteristic function measurement. In our scheme, the compressed cat is realized by creating a superposition of compressed coherent states instead of applying the compression on a conventional cat state. While our technique is capable of both approaches, we chose the former in this work, as it is more technically favorable for the coherent parameters of our system~\cite{pan2022_supp}. In practice, these options for implementing compressed cat states are equivalent up to a different displacement amplitude, which arises from the commutation of squeezing and displacement operators,
$\hat{D}(\alpha) \hat{S}(z)=\hat{S}(z) \hat{S}^{\dagger}(z) \hat{D}(\alpha) \hat{S}(z)=\hat{S}(z) \hat{D}(\gamma)$,
with $\gamma=\alpha \cosh r+\alpha^* e^{i \theta} \sinh r$, and  $\theta$ is phase. We account for this displacement adjustment in our protocol to ensure that the resulting states are equivalent to a squeezed cat state with amplitude $\gamma$.

With this simple technique, we are able to create compressed cat states of size $|\alpha| =1.8, \,\xi\approx \text{-}6.7\,$dB and the desired photon number parity. The real parts of the measured characteristic function for cat (top) and compressed cat states (bottom) of different parities are shown in Fig.~\ref{fig:cat_sqcat}. The effect of the phase-space compression is indicated by the concentration of the interference blobs closer to the origin and their elongation in the opposite direction. This elongation does not influence quantum non-Gaussian features of sub-Planck oscillations, but notably, could impair the parity preservation at high compression levels~\cite{pan2022_supp}. For these states, the centre fringes of the characteristic function indicate the mixture of the coherent amplitudes $\pm\alpha$, with the origin point being always positive. More rapid oscillations in these fringes correspond to larger amplitudes of the cat state. The presence of two blobs along $\mathrm{Re}[\nu]$ directly represents quantum non-Gaussian interference. This cannot be mimicked by semi-classical interference, where the underlying state is a mixture of pure Gaussian states distributed along $\mathrm{Re}[\beta]$ in Wigner function. Moreover, by implementing one dimensional Fourier transformation along $\mathrm{Im}[\nu]=0$, we can directly obtain the marginal distribution of Wigner function along $\mathrm{Im}[\beta]$ corresponding to momentum statistics, and witness high-frequency components corresponding to sub-Planck structure in phase space, which is an intrinsically quantum non-Gaussian attribute caused by the interference~\cite{zurek2001_sub}. In addition, we extract parity $\langle\hat{\mathcal{P}}\rangle= \text{-}0.6 \pm 0.02$ for the odd cat states, by integrating over the full characteristic function. Therefore, it is evident that non-zero values in these interference blobs in the characteristic function representation necessarily corresponds to quantum non-Gaussian interference.

We herein use direct measurements of the interference blob amplitude in the characteristic function to quantify the quantum non-Gaussianity of our states. To investigate the effects of phase-space compression against photon loss, we monitor the decay of the blob amplitude for an odd cat state with different degrees of compression over a duration comparable to the cavity single-photon lifetime. For each point in time, we measure a 1D cut of the real part of the characteristic functions to extract the amplitude of the interference blobs. The results are shown in Fig.~\ref{fig:time_dynamics}(a), along with simulated theoretical curves, for the conventional cat state and three compressed cat states. For the conventional cat state, photon loss leads to an exponential reduction of the blob amplitude from  its maximum value to the noise floor of the measurement, with a time constant of $\tau = 42\pm 2\,\mu$s, as shown in Fig.~\ref{fig:time_dynamics}(a). 

In addition, three snapshots of the characteristic functions are shown in Fig.~\ref{fig:time_dynamics}(b) for the cat and compressed cat states, respectively. It is particularly striking that while the interference blobs of the conventional cat states vanish completely at 100\,$\mu$s, they remain rather prominent in the compressed state with approximately $\text{-}6.7$\,dB compression. Furthermore, the compressed cat states preserve their phase-space distribution without any notable distortions throughout their evolution, which is a key requirement for utilizing cat states for information encoding and highlights the crucial advantage of the low anharmonicity regime we operate in.

Overall, the decay of the quantum non-Gaussian interference is slowed down appreciably to $\tau \approx 87 \pm 5\,\mu$s, $145 \pm 6\,\mu$s, and $147 \pm 8\,\mu$s for the states with $\text{-}3$\,dB, $\text{-}6.7$\,dB, and $\text{-}7.6$\,dB compression, respectively. Our results closely follows the theoretical predictions based on single photon loss, at a timescale of $260\,\mu$s, being the dominant decoherence channel in the system~\cite{pan2022_supp}. Furthermore, we also extract the parity of each state over time by integrating over the full characteristic function, which shows excellent agreement with the timescales obtained by the direct measurements from characteristic functions. 

In summary, we have showcased a versatile technique to perform phase-space compression on cat states in cQED with repeated applications of ECD gates together with single qubit rotations, both of which are readily accessible in standard cQED hardware. We created various compressed cat states based on this technique and sculpted their quantum non-Gaussian features in phase space to be more optimal for the dominant loss mechanism in the system, which is photon loss. We then probed their intrinsically quantum mechanical features by directly measuring specific regions in their characteristic function. We demonstrated that a more compact landscape in phase-space significantly enhances the protection of the compressed cat state's quantum non-Gaussian interference features against photon loss. 

Our study brings forth valuable insights for both intrinsic dynamics of CV quantum non-Gaussian resource states and useful applications built upon them. Fundamentally, our results demonstrate that the general strategy of reshaping the phase space of quantum states, using optimized ECD operations~\cite{park2022_slowing}, effectively preserves their quantum features and enhances their resilient against losses in the hardware. The specific technique to compress the cat state provides a versatile tool for storing and protecting useful non-Gaussian resource states in highly-linear bosonic modes in cQED, and affords a flexible test-bed for exploring fundamental physics~\cite{haroche2013_nobel} and CV-based quantum metrology~\cite{munro2002weak, joo2011quantum, facon2016_sensitive, knott2016practical, duivenvoorden2017_single}. Specifically, as these compressed cat states have well-defined parity, they are natural candidates for efficient quantum error correction in bosonic modes, where parity is often used as the main error syndrome. They are inherently more robust codewords compared to conventional cat states, thanks to their compact phase-space distribution and the resulting protection of their quantum non-Gaussian againt decay photon-loss. This is further investigated in two recent theory proposals~\cite{xu2022_autonomous, hillmann2022_quantum} indicating that squeezed cat states, together with manifold stablization, offers a promising path towards protected logical qubits encoded in superconducting cavities. Our work marks a significant step towards experimentally realising these new paradigms for quantum error correction and fault-tolerant quantum computing. 

\textbf{Acknowledgements}
We thank Yifan Li, Steven Touzard, and Berge Englert for the fruitful discussions related to this work. The quantum amplifier used in this experiment is graciously provided by Dr.~Ioan Pop and Dr.~Patrick Winkel from Karlsruhe Institute of Technology. This research is supported by the National Research Foundation, Singapore, and the Ministry of Education, Singapore under the Research Centres of Excellence Programme. R.F. acknowledges funding from the project LTAUSA19099 of the Czech Ministry of Education, Youth and Sports (MEYS CR) and the European Union’s 2020 research and innovation programme (CSA Coordination and support action, H2020-WIDESPREAD-2020-5) under grant agreement No.\,951737 (NONGAUSS). Y.Y.G.~acknowledges the support of the National Research Foundation Fellowship (NRFF12-2020-0063) and the Ministry of Education (Grant ID 21-0054-P0001), Singapore. 
\\ \\ 
\bibliography{reference_main.bib}    

\clearpage

\maketitle

\section{Device parameters}

\begin{figure}
    \centering
    \includegraphics[scale=0.5]{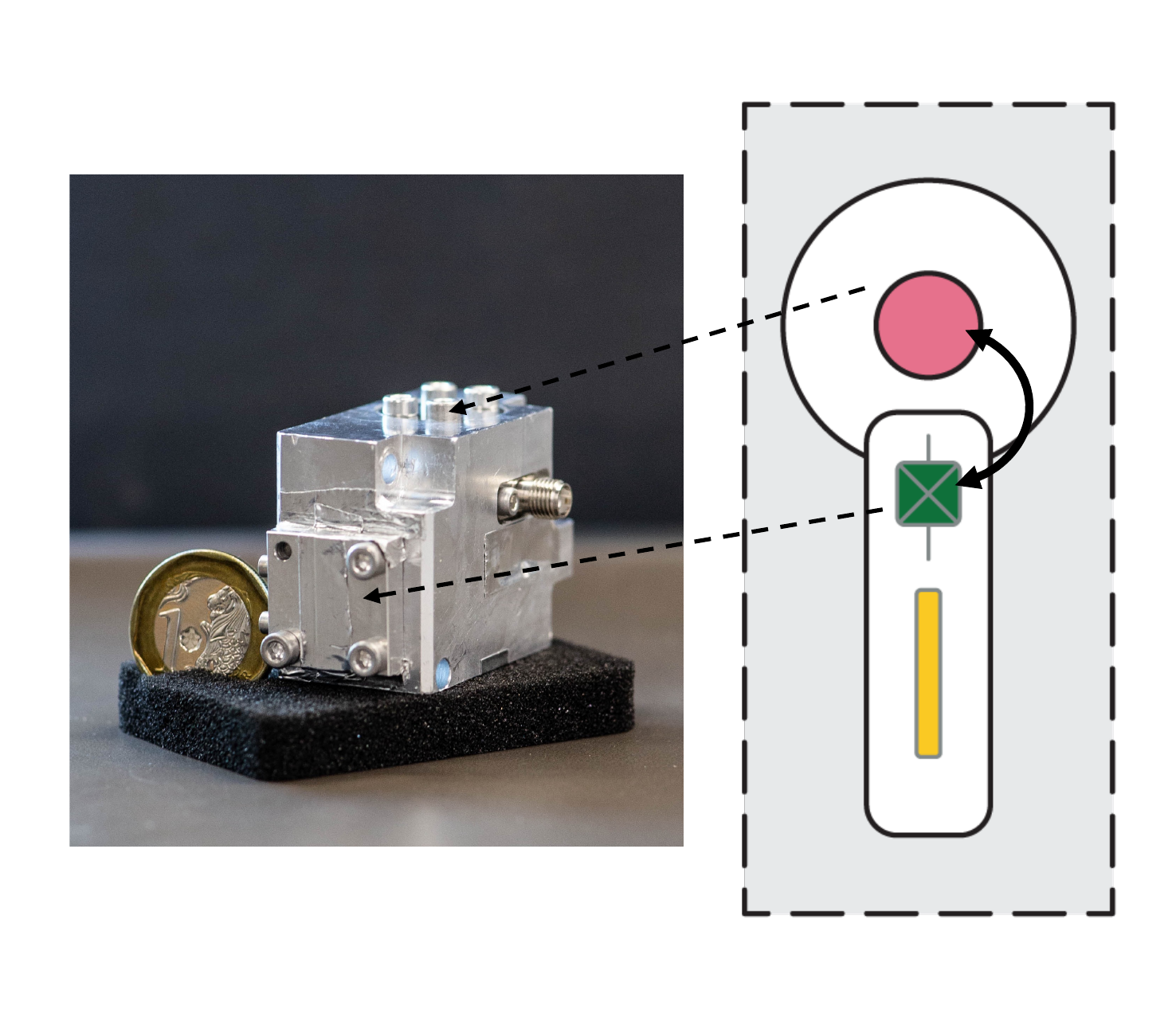}
    \caption{\textbf{cQED hardware.} Photograph of a device machined out of high-purity aluminum consisting of one storage cavity (pink), single junction transmon (green), and planar readout resonator (gold). }
    \label{sfig:device}
\end{figure}

Our device consists of a three-dimensional (3D) superconducting microwave cavity, an ancillary transmon, and a planar readout resonator as shown in Fig.~\ref{sfig:device}. The cavity is machined out of high-purity (4N) aluminum and is considered as a 3D version of a $\lambda/4$ transmission line resonator between  a centre stub and cylindrical wall. The readout resonator deposited together with the transmon on a sapphire chip using double-angle evaporation. The sapphire chip is inserted into the tunnel, with the  pads of the transmon slightly extending into the coaxial cavities to provide the capacitive coupling. Here, we design a weak dispersive coupling to ensure minimal non-linearity in the cavity mode, such that it serves as a highly harmonic quantum memory that can store multi-photon bosonic states with negligible distortion. The ancillary transmon provides the capability of fast control of the single cavity. The low-Q readout resonator, together with a quantum-limited amplifier, allows fast single-shot measurement.

\begin{table}[h]
    \centering
    \begin{tabular}{c|c|c|c|c}
    
    \hline
    \hline

    & Frequency & $\chi$ to transmon & $\chi$ to cav & $\chi$ to RO\\
    \hline
    transmon & 5.1461\,GHz & 205.4\,MHz & 80\,kHz & 1\,MHz  \\
    cavity   & 6.5428\,GHz  & 80\,kHz  & $\sim$10\,Hz   & --  \\
    RO  & 7.4418\,GHz  & $\sim1$\,MHz   & --  & -- \\
         \hline
    \end{tabular}
    \caption{\textbf{Hamiltonian parameters} Summary of the key system parameters. The non-linearity of the cavity is below our measurement sensitivity, and is extracted based on the simulation of the system. RO: readout, cav: cavity. }  
    \label{table:hamiltonian}
\end{table}

With the weak dispersive coupling between the transmon and the cavity, we are no longer able to extract $\chi$ by standard number-splitting measurements. Instead, we derive it by measuring the rotation of coherent state in characteristic function space over time. Experimentally, we generate a large coherent state with transmon in excited state, and allow it to evolve over a variable amount of time. This causes the coherent state to rotate at a rate governed by the dispersive coupling term $\chi\hat{a}^{\dagger}\hat{a}/2$. We extract the resulting rotation angles by fitting the 2D characteristic function of coherent state at different evolution times $\Delta t$ to obtain $\chi$ from \textbf{$\Delta\theta = \chi \Delta t$}. The Hamiltonian parameters of the system are summarized in Table~\ref{table:hamiltonian}.

The cavity energy relaxation time $T^\mathrm{c}_1$ is obtained by measuring a decay of coherent state over time, as discussed in Ref.~\cite{reagor2015_thesis}. We generate a large coherent state and this state falls to the origin with a characteristic rate $\kappa$, which is probed by applying a highly selective transmon $\pi$ pulse conditioned on $|0\rangle$ of cavity to measure the population of the vacuum $P_0$. 

For the cavity $T_2$, we displace the cavity by $\beta$ and measure its characteristic function under different decay times as discussed in the supplementary materials of Ref.~\cite{Camp2020_quantum}. Through comparing the measured result with the simulation which assigned by different $T_2$ to obtain the highest fidelity, we obtain a bound on $T^\mathrm{c}_2$ to be $\geq$\,5\,ms. This is limited by the residual thermal population of the transmon at $\sim1.5\%$. The coherence parameters of the system are summarized in Table~\ref{table:coherence}.

\begin{table}[hbt!]
    \centering
    \begin{tabular}{c|c|c|c}
    \hline
    \hline
                & $T_1 (\mu$s) & $T_2 (\mu$s) & $T_{2\mathrm{e}} (\mu$s)\\
        \hline
         transmon & 20 & 18 & 20 \\
         cavity & 260 & $\geq$\,5000 & -- \\
         \hline
    \end{tabular}
    \caption{\textbf{Coherence parameters} Summary of typical coherence time scale of ancillary transmon and storage cavity, respectively. }  
    \label{table:coherence}
\end{table}

\section{Characteristic function measurements}    
The characteristic function is defined as ${C(\nu)=\langle D(\nu) \rangle}$. It can be measured directly using conditional displacement operations and single transmon rotations as discussed in Ref.~\cite{Camp2020_quantum}. The conditional displacement gate is first calibrated by the measuring the characteristic function of vacuum state to get a unit displacement amplitude. Experimentally, we do so by adjusting the scaling of the conditional displacements by sweeping amplitude of the displacement gates so that the measured characteristic function of the vacuum state is a Gaussian with standard deviation of $\sigma = 1$. To calibrate the unconditional displacement gate, we can use the forward-and-back sequences~\cite{Eick2021_fast} consisting of $D(-i\alpha)CD(-1)D(i\alpha)CD(1)$ with the cavity initially in the vacuum state and transmon in $|g\rangle+|e\rangle$. This procedure displaces the cavity to a coherent state first and then back to the vacuum state in the end such that a geometrical phase is accumulated. We measure the expectation value of $\sigma_x$ by varying $\alpha$ to find the appropriate amplitude scaling for a unconditional displacement $\alpha=1$, which corresponds to an oscillation with period of $1/\pi$. 


\section{Comparison of different metrics for non-classicality}

Here, we compare the methods of extracting the quantum interference features of the cat and compressed cat states of amplitude $\alpha=1.8$. More specifically, we analyse the corresponding Wigner functions, sub-Planck structures and fidelities from reconstructed density matrices and direct integral of the characteristic function over the phase space. 

\subsection{Wigner function reconstruction}
\begin{figure}[htb!]
\includegraphics[scale=1]{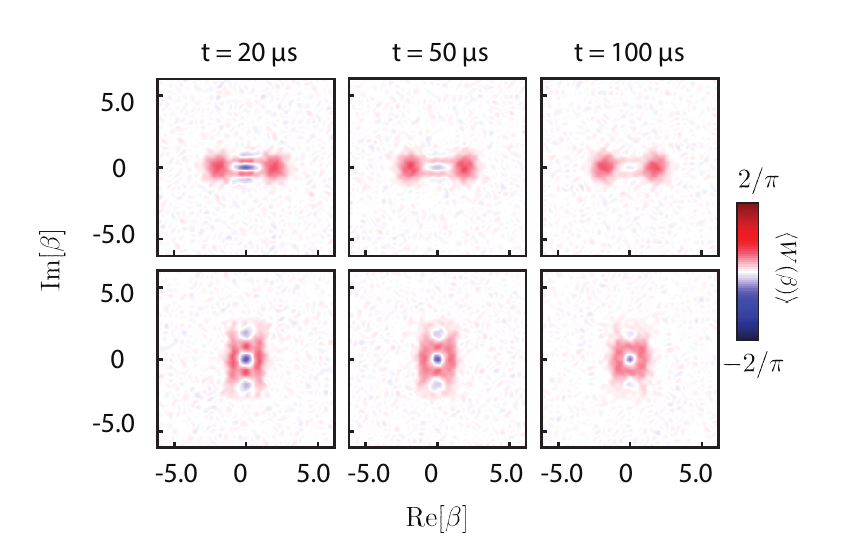}
\caption{\textbf{Discrete Fourier transformation}  Wigner function of odd cat with 0\,dB. (top) and -7.6\,dB. (bottom) squeezing at the decay time of $20\,\mu$s, $50\,\mu$s and $100\mu$s respectively. }
\label{sfig:FFT_wigner_from_char} 
\end{figure}

We obtain the Wigner function, ${W}{(\beta)}$, from the characteristic function, $C(\nu)$, by performing two-dimensional discrete Fourier transform of $C(\nu)$ based on the following equation:
\begin{equation}
     {W}{(\beta)} = \frac{1}{\pi^2}\int C(\nu)e^{\beta\nu^*-\beta^*\nu}d^2\nu.
\end{equation} \label{seq:wigner_from_char}  
Prior to the discrete Fourier transform (DFT), we pad the raw characteristic function data with zeros outside the measurement range. This allows a more accurate transformation between the two representations by extending phase-space beyond the measurement range. The resulting Wigner functions for an odd cat state and a $-7.6$\,dB. compressed odd cat state at three different decay times are shown in Fig.~\ref{sfig:FFT_wigner_from_char}. These Wigner functions closely resemble the expected behaviors as shown in the main text, illustrated by the direct characteristic measurements. We observe the quantum interference features, denoted here by the presence of Wigner negativity, are substantially more pronounced in the compressed cat state after 100\,$\mu$s decay time compared to that of the cat state.

Furthermore, we can directly obtain the photon number parity of odd cat states from the value of the Wigner functions at the origin of the phase space $W(0,0)$. Their time dynamics (Fig.\,4(a) inset in main text) closely echoes that of the decay of interference blob amplitudes extracted directly from the characteristic functions. We observe that with phase-space compression, the parity decays significantly more slowly in the presence of photon loss while that of the conventional cat state diminishes to zero rapidly. 


\subsection{Sub-Planck phase space structure}
Another useful tool to witness non-classical interference is the presence of high-frequency components corresponding to sub-Planck structures in the phase space~\cite{zurek2001_sub}. In order to extract this feature from the measured characteristic functions, we implement one dimensional Fourier transformation along $\mathrm{Im}[\nu]=0$ and compare to that of vacuum.

\begin{figure}[h!]
    \centering
    \includegraphics[scale=1]{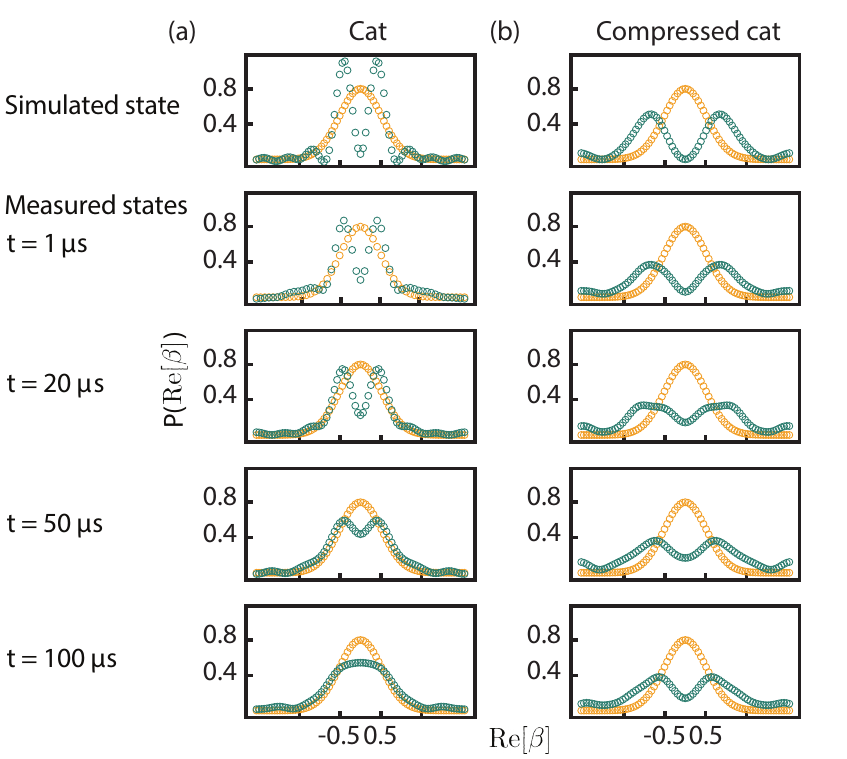}
    \caption{\textbf{Decay of sub-Planck structure} We implement a one dimensional Fourier transformation along $\mathrm{Im}[\nu]=0$ of the ideal (top row) and measured characteristic function. We obtain the marginal momentum distribution of $\mathrm{Im}[\beta]$ for (a) cat (left, green) and (b) -7.6\,dB compressed cat (right, green) of different decay times, respectively, and compare the width of resulting oscillatory features to that of ideal vacuum (gold).}
    \label{sfig:subplanck}

\end{figure}

Here, we illustrate the behavior of the cat and a -7.6\,dB compressed cat state at times $1\,\mu$s, $20\,\mu$s, $50\,\mu$s, and $100\,\mu$s. For the compressed cat (Fig.~\ref{sfig:subplanck}(b)), the non-classical sub-Planck structures are still notably present at $100~\mu$s. However, for the cat states (Fig.~\ref{sfig:subplanck}(a)), the sub-Planck features vanish rapidly and are no longer visible at $100~\mu$s. From these results, we again confirm that the quantum non-Gaussian features of compressed cat states are significantly better preserved compared to conventional cat states.

\subsection{Density matrix reconstruction} 
We reconstruct the density matrices of the states using convex optimization~\cite{strandberg2022_simple} with the measured characteristic functions as the input. We experimentally verify that the imaginary components of the characteristic functions are negligible (i.e. within measurement noise) and perform reconstruction using the real parts for an odd cat state and a compressed cat state at decay times $1\,\mu$s, $20\,\mu$s, $50\,\mu$s, and $100\,\mu$s with 1000 averages per point each. The Hilbert space size 20 chosen for the reconstruction protocol is based on simulations of photon number distributions of the ideal cat states with the same amplitude. The fidelities $\mathcal{F}_\mathrm{cat, cv}$ ($\mathcal{F}_\mathrm{ccat, cv}$), defined as the overlap of the ideal cat (compressed cat) state and the experimentally reconstructed density matrices, are shown in Table.~\ref{stable:fidelity}. The associated uncertainties in these values are computed using standard bootstrapping techniques.

It is again, apparent that $\mathcal{F}_\mathrm{cat, cv}$ decay more rapidly compare to $\mathcal{F}_\mathrm{ccat, cv}$. For instance, as the decay time increases from $1\,\mu$s to $20\,\mu$s, the fidelity of cat state decreased sharply from $75\%$ to $55\%$. In contrast, for the compressed cat state, the fidelity dropped notably more slowly from $61\%$ to $56\%$.

\subsection{Direct integration over phase-space} 
An alternative method to calculate the fidelity of states of interest is directly integrating the characteristic function over the phase space. For a state with characteristic function $C_{\mathrm{exp}}{(\nu)}$, we compute this overlap to the ideal target state $C_{\mathrm{ideal}}{(\nu)}$ using the following equation:
\begin{equation}
     {\mathcal{F}_\mathrm{int}} = \frac{1}{\pi}\int C_{\mathrm{ideal}}{(\nu)}C_{\mathrm{exp}}{(\nu)}^*d^2\nu.
\end{equation} 
The calculated fidelities of cats $\mathcal{F}_\mathrm{cat,int}$ and compressed cats $\mathcal{F}_\mathrm{ccat,int}$ at different decay times are presented in Table.~\ref{stable:fidelity}. It is consistent with the fidelities obtained from density matrix reconstruction, where the compressed cat state retains its fidelity much better in the presence of photon loss. 

\begin{table}[h!]
\begin{tabular}{c|cccc}
\hline
\hline
\textrm{State}&
\textrm{$\mathcal{F}_\mathrm{cat,cv}$~$(\%)$} &
\textrm{$\mathcal{F}_\mathrm{cat,int}$~$(\%)$}&
\textrm{$\mathcal{F}_\mathrm{ccat,cv}$~$(\%)$}& 
\textrm{$\mathcal{F}_\mathrm{ccat,int}$~$(\%)$}\\
\colrule
1 $\mu$s & 75$~(\pm 0.5)$ & 74.8$~(\pm 0.6)$ & 61$~(\pm 0.7)$ & 67.1$~(\pm 0.7)$\\
20 $\mu$s & 55$~(\pm 0.6)$ & 61.52$~(\pm 0.8)$ & 56$~(\pm 0.5)$ & 60.4$~(\pm 0.7)$\\
50 $\mu$s & 45$~(\pm 0.7)$ & 47.88$~(\pm 0.7)$ & 50$~(\pm 0.6)$ & 53.9$~(\pm 0.6)$\\
100 $\mu$s & 40$~(\pm 0.7)$ & 41.15$~(\pm 0.7)$ & 43$~(\pm 0.5)$ & 46.4$~(\pm 0.7)$\\
\hline
\end{tabular}
\caption{Comparison of fidelities extracted from density matrix reconstruction and direct integration of characteristic function.}
\label{stable:fidelity}
\end{table}

\section{Ideal compression for parity protection}
Photon loss acts as a low-pass filter with a 2D Gaussian profile in the phase space. For the preservation of interference features of cat states, the more strongly compressed (i.e. closer in to the origin in phase space), the more pronounced the protection is. However, for the photon number parity, which is an important observable used in quantum error correction schemes, the resulting elongation in the opposite quadrature due to strong compression in phase space causes the cat states to exceed the Gaussian filter. 

\begin{figure}[htb!]
    \centering
    \includegraphics[scale=1]{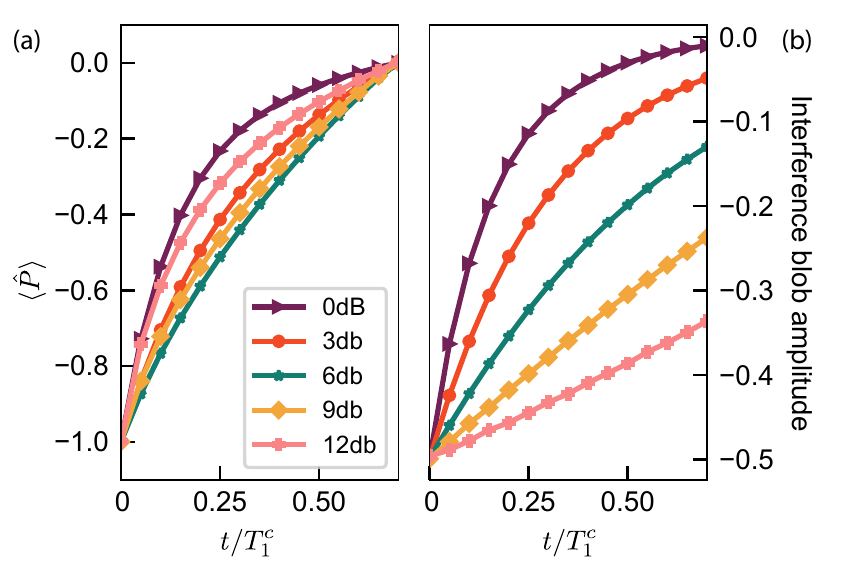}
    \caption{\textbf{Simulation result of decay of photon number parity and interference blob amplitude}.
    Parity decay (a) and interference blob decay in the characteristic function (b) for cat states with different degrees of compression. 
    While there exists an optimal compression level to protect parity, the decay of interference blob is consistently slower as compression increases.} 
    \label{sfig:optimal_compression}
\end{figure}

Here, we investigate this effect by simulating the behavior of the interference blobs and the photon number parity of a cat state under different compression levels as shown in Fig.~\ref{sfig:optimal_compression}. In the case of the interference blob amplitude, the decay becomes consistently slower as we increase the compression in phase space. In comparison, the reduction in parity is mitigated by the compression up to -6\,dB. Further compression leads to an acceleration in its decay, due to the extension in the opposite direction in phase space beyond the Gaussian filter. As the photon number parity requires integration over the entire characteristic function, it is sensitive to the elongation of the state under compression. Therefore, for the preservation of parity, there exists an optimal compression level that is determined by the photon loss rate in the system and the size of the cat state. 

\section{Compressed vacuum creation}

The compressed vacuum states are generated via a deterministic protocol, as proposed in Ref.~\cite{Hastrup2021_Unconditional}. It consists of multiple repetitions of the two unitary gates $\hat{U}_k$ = exp$(iu_k\hat{P}\hat{\sigma}_x)$ and $\hat{V}_k$ = exp$(iv_k\hat{X}\hat{\sigma}_y)$ with $\hat{X} = (\hat{a} + \hat{a}^\dagger)/2$ and $\hat{P} = i(\hat{a}^\dagger - \hat{a})/2$. The unitaries $\hat{U}_k$ and $\hat{V}_k$ are essentially conditional displacement in opposite directions in phase space, sandwiched by transmon rotations. 

The key to generate compressed vacuum states is to choose the ideal number of repetitions along with numerically optimized interaction coefficients $u_k$ and $v_k$. While more repetitions translate into higher achievable compression in theory, the increased gate time has to be balanced against the dominant decoherence timescale. In our case, the transmon decoherence $T_{2\mathrm{e}}$ is the main limiting factor. For the results presented in this paper, we chose three repetitions, with each $\hat{U}\hat{V}$ step taking $\sim1.37\,\mu$s.

In the proposed protocol, the more repetitions of the $\hat{U}\hat{V}$ operations are applied, the closer the resulting state would be to a genuinely squeezed vacuum. However, in the limit of only three repetitions, the final states are only approximately squeezed, with some outlying non-Gaussian features in the phase space. However, these states are effective compressed in the desired quadrature, with a reduction in their distribution consistent with up to -7.6\,dB of compression. Therefore, the presence of non-perfect squeezing is not a hindrance to our scheme, as the protection of the quantum interference features arises from the state being more compact in phase space and is not dependent on the degree of genuine squeezing. 

To generate the coefficients $u_k, v_k$, we numerically optimize them such that the final state has maximal overlap with a target squeezed state, with additional cost parameters given by the physical constraints in our system, such as the maximally achievable ECD displacement for a given gate time. In simulation, the states generated by the protocol have overlaps of ${\mathcal{F}>\,0.99}$ with respect to the target state. The optimized parameters for target squeezed states with -3, -5, -6 and -7\,dB can be found in Table~\ref{interaction_params}.

\begin{table}[hbt!]
\begin{tabular}{ c|c c c c c c } 
\hline
\hline
 & $u_1$ & $v_1$ & $u_2$ & $v_2$ & $u_3$ & $v_3$  \\
\hline
-3\,dB& 1.39 & 0.51  & -0.2 & -0.46 & -0.32 & -0.65 \\
-5\,dB& -0.48& 0.51& -1.85& -0.31& 0.56& 0.91  \\
-6\,dB& 1.6&  0.39& -0.48& -1.04& -1.11& 0.32 \\
-7\,dB& -0.83&  0.56&  1.3& -0.56& -1.26& 0.39  \\
\hline
\end{tabular}
\caption{\textbf{Summary of optimised coefficients}}
\label{interaction_params}
\end{table}

The corresponding quadrature compression values in dB in Table~\ref{quad_squeezing}. They indicate that three repetitions of $\hat{U}\hat{V}$ allow the creation compressed vacuum states with compression close to the corresponding ideal squeezed state. The experimental compression values are extracted by Gaussian fits to 1D cuts along each quadrature. We then calculate the level of compression by $20\log_{10}[\sigma/\sigma_\mathrm{vac}]$.

\begin{table}[hbt!]
\begin{tabular}{ c | c c | c c } 
\hline
\hline
 &\multicolumn{2}{c|}{Theory} & \multicolumn{2}{c}{Exp.}\\
\hline
 & $\langle\hat{P}^2\rangle$  & $\langle\hat{X}^2\rangle$  & compres. in $\hat{P}$ &  compres. in $\hat{X}$  \\
\hline
-3\,dB& 2.96 &  -2.98  & 2.6 & -3  \\
-6\,dB& 5.71& -5.93& 5.4& -6.7 \\
-7\,dB& 5.9&  -7.24&  6.4& -7.6 \\
\hline
\end{tabular}
\caption{\textbf{Summary of achieved compression on vacuum state}}
\label{quad_squeezing}
\end{table}

\begin{figure}[h!]
\includegraphics{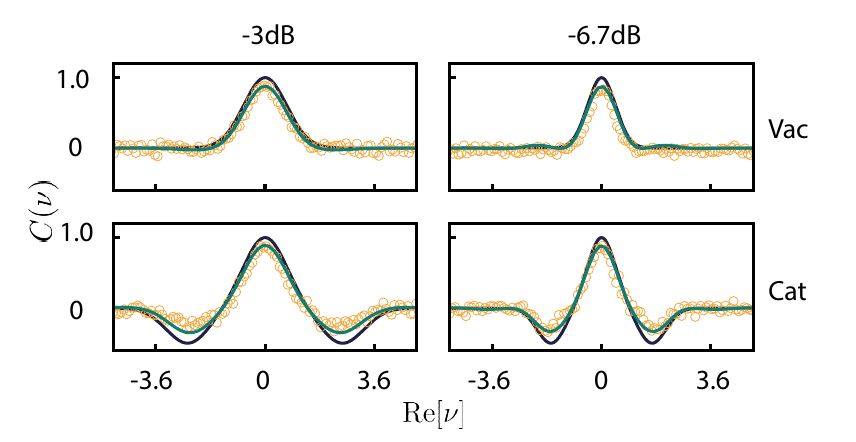}
\label{sfig:suppl_squeezing}
\caption{\textbf{Comparison of simulated and measured characteristic functions.} Black lines (Green lines) indicate simulated characteristic function of compressed states without loss (with transmon $T_{\mathrm{2e}}$). Gold circles denote experimental data. The simulation with loss is scaled with the data contrast. The appearance of side loops is clearly visible, and consistent between data and simulated behavior. } 
\end{figure}

It is important to note that the optimization process does not take into account the noise model of the hardware. Therefore, due to the decoherence of the transmon, we observe a slight the deviation from the target state. To verify this, we show the real part of 1D characteristic function cuts of the compressed vacuum states (-3\,dB and -6.7\,dB) in Fig.~\ref{sfig:suppl_squeezing}. The compression shows in smaller $\sigma$ in characteristic function. The appearance of side loops can clearly be seen in the simulation and data of the -6.7\,dB state, reducing the amount of real squeezing along the cut. The simulation curves are scaled using the measured char functions origin point. 

\section{Commuting compression and cat state creation}
Here, we would like to illustrate the flexibility of the phase-space manipulation technique demonstrated in this work. While we chose to create a superposition of two compressed coherent states, we can also first create a cat state and then apply the compression operation subsequently. The protocol largely remains the same. We initialize the cavity in a cat state and numerically optimize the parameters such that the overlap is maximal with the desired compressed cat state. We simulate this process to verify its practical viability. Using three $\hat{U}\hat{V}$ steps of the $\hat{U}\hat{V}$ operations, we can (in simulation) achieve -3, -5, -6 and -7\,dB compression with $\mathcal{F}>0.99$.

The specific choice of our implementation is made based on practical optimizations for the device's coherence parameters. As a compressed vacuum spans a smaller range of energy levels, our chosen protocol reduces as the exposure additional cavity decoherence during state creation process. This can be readily modified or adapted to bring more optimal implementations on cQED hardware with different system parameters.  

\section{Error budget}
The fidelities of the compressed cat states we created in this work, extracted using the different methods mentioned in Sec.III, are generally in the range of $65(\pm 5)\%$. This is largely limited by the decoherence timescales in our hardware. To analyse this, we chose the action of a single $\hat{U}$ gate as the base for gauging the imperfections involved in our state creation and measurement process, which are essentially $\hat{U}$ or $\hat{V}$ operations sandwiched between single transmon rotations. The fidelity's are calculated as the overlap of the ideal state with the overlap of the ideal suffering from a single decoherence mechanism $ {\mathcal{F} = \langle\psi_\mathrm{ideal}|\rho_\mathrm{loss}|\psi_\mathrm{ideal}\rangle}$. The simulated infidelities due to the different sources of non-idealities in the device are shown in Table~\ref{tab:infidelities}. They are calculated via master equation simulations including only the respective decoherence mechanism.\\

\begin{table}[hbt!]
\begin{tabular}{c|c}
\hline
\hline
Error channel \quad & Estimated infidelity \\
\hline
ancilla dephasing &  4\% \\
ancilla decay &  2\% \\
cavity dephasing &  1\% \\
cavity decay &  0.01\% \\
$\hat{U}$ & $\approx$ 7 \% \\
readout $|g\rangle\rightarrow |g\rangle$ & 1.4\% \\
readout $|e\rangle\rightarrow|e\rangle$ & 5\% \\
\hline
\end{tabular}
\label{tab:infidelities}
\caption{\textbf{Error budget} Estimated infidelities due to SPAM errors to the reduction in fidelities of our cats and compressed cat states. Ancilla dephasing and decay together with readout errors are the dominant sources of error in our system. }
\end{table}

First, let us compare this proposed error budget with experimentally observed imperfections in our characteristic function measurement. In our results, the measured vacuum state shows a maximum contrast of $\approx 88.4\%$. The implementation requires a single $\hat{U}$-type gate and a transmon readout. Summing up the individual contributions, we expect the $\hat{U}$ operation to have a fidelity $\approx 7\%$ and the readout fidelity of $(P_\mathrm{ee} + P_\mathrm{gg})/2 \approx 3.2\%$. This results in an overall limit on the measurement fidelity to $\approx 90\%$, which is consistent with the contrast of the measured vacuum state. We use this as a normalisation factor for the subsequent data of the cat and compressed cat state creation to isolate the state preparation errors from that of the measurement. 

For the compressed vacuum and cat states shown in the main text, the state creation process involves 3 sets of $\hat{U}\hat{V}$ operations and one measurement to project the transmon state. The post-selection process after this measurement effectively removes the contribution of transmon $T_1$, making the fidelity of each $\hat{U}$ or $\hat{V}$ gate approximately $5\%$. Therefore, using the simulated error budgets, we expect them to suffer from $\approx 33\%$ infidelity, which is again consistent with the data presented. 

While decoherence is the dominant limitation, other mechanisms such as the Kerr effect, calibration inaccuracies, and residual imaginary components in the characteristic functions, etc. could also introduce some imperfections to states we consider here. However, as these are small compared to decoherence errors, we do not have the resolution to analyse them in detail. Overall, if this protocol is applied on a system with better coherence parameters, these investigating and minimising these imperfections will become more crucial. 


\end{document}